\documentclass[12pt]{article}
\usepackage[textures]{graphics}

\parskip=\bigskipamount
\newcommand{\beq}{\begin{equation}}
\newcommand{\eeq}{\end{equation}}
\newcommand{\beqa}{\begin{eqnarray}}
\newcommand{\eeqa}{\end{eqnarray}}
\newcommand{\ket} [1] {\vert #1 \rangle}
\newcommand{\bra} [1] {\langle #1 \vert}

\newcommand{\proj}[1]{\ket{#1}\bra{#1}}

\newcommand{\mod}{~{\rm mod}~}

\begin{document}

\title{{\sc Security of Quantum Key Distribution with entangled Qutrits.}}

\author{Thomas Durt$^1$, Nicolas J. Cerf$^{2}$,  Nicolas Gisin$^3$ and  M. \.Zukowski$^4$\\
\\
$^1$ Toegepaste Natuurkunde en Fotonica, Theoretische Natuurkunde,\\
Vrije Universiteit Brussel, Pleinlaan 2, 1050 Brussels, Belgium\\
$^2$ Ecole Polytechnique, CP 165, Universit\'e Libre de Bruxelles,\\
1050 Brussels, Belgium\\
$^3$ Group of Applied Physics - Optique, Universit\'e de Gen\`eve,\\
20 rue de l'Ecole de M\'edecine, Gen\`eve 4, Switzerland, \\
$^4$Instytut Fizyki Teoretycznej 
i Astrofizyki Uniwersytet, \\
Gda\'nski, PL-80-952 Gda\'nsk, Poland}

\date{}

\maketitle

\begin{abstract} The study of quantum cryptography and quantum entanglement
have traditionally been based on two-level quantum systems (qubits). In this
paper, we consider a generalization of Ekert's entanglement-based quantum
cryptographic protocol where qubits are replaced by three-level systems
(qutrits).  In order to investigate the security against the optimal
individual attack, we derive the information gained by a potential
eavesdropper applying a cloning-based attack. We exhibit the explicit
form of this cloner, which is distinct from the previously known cloners,
and conclude that the protocol is more robust than the ones based
on entangled qubits as well as unentangled qutrits.
\end{abstract}


\section{Introduction}

Quantum cryptography aims at distributing a random key in such a way that the
presence of an eavesdropper who monitors the quantum communication
is revealed via the induced disturbances in the transmission of the key
(for a review, see e.g. \cite{RMP}). 
Practically, in order to realize a cryptographic protocol, it is
enough that the key signal is encoded into quantum states that belong to
incompatible bases, as in the original protocol of Bennett and Brassard
known as BB84\cite{BB84}.
In 1991, Ekert suggested to base the security of quantum
cryptography on properties of the maximally entangled two-qubit state
or EPR state\cite{Ekert}. The key signals are derived from measurements 
when they lead to perfect correlations (same base used by the two parties), 
and otherwise data for a Bell \cite{bell} or 
Clauser-Horne-Shimony-Holt (CHSH) \cite{chsh} inequality test 
are collected and  used to reveal the presence of an eavesdropper.
Recently,  it was shown that the violation of Bell-type inequalities is more
pronounced in the case of entangled qutrits (i.e., 3-dimensional systems)
than entangled qubits \cite{zukozeil,chenzuko,collins}. Also, several
qutrit-based cryptographic protocols were shown to be more secure than their
qubit-based counterparts\cite{HBech-APeres,bourennane,boure2,bruss3D}.
It appears therefore very tempting to investigate
the performances of a generalization of Ekert's protocol relying on
a pair of entangled qutrits \cite{ZZH} instead of qubits.
\par

>From the experimental viewpoint, there are several ways of physically
realizing qutrits using photons. The first possibility 
is to utilize multiport-beamsplitters, and more specifically those that
split the incoming single light beam into three \cite{ZZH}.
The second one exploits the polarization degree of freedom. 
However, since this is intrinsically
a two-dimensional variable, one needs to use two photons 
per qutrit \cite{Howell,Burlakov}. A third possibility, which
uses only one photon per qutrit, exploits the spatial angular momentum
of photons \cite{Mair}. Finally, another realization of qutrits, possibly
the most straightforward one, exploits time-bins \cite{Tittel}. 
This approach has already been demonstrated for entangled photons 
up to eleven dimensions \cite{DeRiedmatten}. Thus, exploring 
an entanglement-based quantum cryptographic protocol that uses qutrits
instead of qubits can lead to new applications of 
quantum informational technology as it lies 
in the reach of the current state-of-the-art quantum optical techniques.
\par

In what follows, we shall analyze the security of this entanglement-based
protocol against individual attacks (where the eavesdropper Eve monitors the
qutrits separately or incoherently). To this end, we will consider a fairly
general class of eavesdropping attacks that are based on (state-dependent)
quantum cloning machines\cite{CERFPRL,CERFA,CERF}.
This will yield an upper bound on the acceptable
error rate, which is a {\em necessary} condition for security against
individual attacks, that is, higher error rates cannot permit to establish
a secret key using one-way communication.
We will show that this maximum acceptable error rate is higher,
with this qutrit protocol, than with Ekert's qubit protocol, and even slightly
higher than with a three-dimensional extension of BB84.

\par

\section{The four qutrit bases that maximize the violation of local
realism}

In the protocol Ekert91\cite{Ekert}, the four qubit bases chosen by Alice and
Bob (the authorized users of the quantum cryptographic channel) are the four
bases that maximize the violation of the CHSH
inequalities \cite{chsh}. They consist of two pairs of mutually unbiased
bases\footnote{By definition, two orthonormal bases of an N-dimensional
Hilbert space are said to be mutually unbiased if the norm of the scalar
product between any two vectors belonging each to one of the bases is equal
to ${1\over \sqrt N}$.}. When representing these four bases on the Bloch
sphere, their eight states form a perfect octagon [see Fig.1
(right)]. Similarly, there exists a natural generalization of this set of
 bases in the case of qutrits \cite{Durt}. In analogy with the CHSH qubit
bases, which belong to a great circle, these four qutrit bases
belong to a set of
bases parametrized by a phase $\phi$ on a generalized equator, which we shall
call the $\phi$-bases from now on. The expression of the component states of
any $\phi$-basis in the computational basis $\{\ket{0},\ket{1},\ket{2}\}$ is
\begin{eqnarray}
\ket{l_{\phi}}&=&{1\over
\sqrt 3}\sum_{k=0}^2 e^{ik({2\pi l\over 3}+\phi)}\ket{k} \nonumber\\
&=&{1\over
\sqrt 3}e^{i({2\pi l\over 3}+\phi)}\left(\ket{1}+\cos({2\pi l\over
3}+\phi)(\ket{0}+\ket{2})+\sin({2\pi l\over 3}+\phi)(-i)(\ket{0}-\ket{2})
\right),
\end{eqnarray} with $l=0,1,2$. Obviously, these basis vectors form an
equilateral triangle on a great circle centered in $\ket{1}$. When
$\phi$ varies, these triangles turn around $\ket{1}$. Note that the state
$\ket{1}$ plays
a privileged role compared with the states $\ket{0}$ and $\ket{2}$. The
invariance under a cyclic permutation of the basis vectors of the
computational basis is indeed broken in the $\phi$-bases because it can happen
that  $k=k' \mod 3$ while
$e^{ik\phi}\not= e^{ik'\phi}$ ($k,k'=0,1,2$) when $\phi \not= {2\pi l\over 3}$
(l=0,1,2). It has been shown that when local observers measure the
correlations exhibited by the maximally entangled state
\begin{equation}
\ket{\phi_3^+}={1\over\sqrt{3}}(\ket{0}\otimes
\ket{0}+\ket{1}\otimes \ket{1}+\ket{2}\otimes
\ket{2})
\end{equation} in the four $\phi$-bases obtained when
$\phi_i\,=\,{2\pi\over 12}\cdot i$ (with $i=0,1,2,3$), then the degree of
non-classicality that characterizes the correlations is higher than the
degree of non-classicality allowed by Cirelson's  theorem \cite{cirelson} for
qubits, and also higher than for a large class of other qutrit bases. This
can be shown by estimating the resistance of the non-classicality of
correlations against noise admixture \cite{zukozeil}, or by considering
generalizations of Bell inequalities to a situation in which trichotomic
observables are considered \cite{chenzuko,collins} instead of dichotomic
ones. Note that the states making up the four qutrit bases which maximize the
violation of local realism (we shall call them the {\it optimal bases}
from now on) form a perfect dodecagon,
which generalizes the octagon encountered in the
qubit case.

Finally, it is worth noting that the state that optimizes the
violation of local realism when considering the four optimal bases
is not the maximally entangled state, but the state
$\ket{\phi_{mv}}={1\over\sqrt{n}}(\ket{0}\otimes
\ket{0}+\gamma\ket{1}\otimes \ket{1}+\ket{2}\otimes
\ket{2})$ where $\gamma= {(\sqrt{11}-\sqrt{3})\over 2}$ 
and $n=2+\gamma^2$ \cite{acin}.  This state is
not invariant under a cyclic permutation of the basis vectors of the
computational
basis. We noted already that this invariance is
broken by the
$\phi$-bases. We shall not discuss here the implementation of this state in
quantum cryptography.

\section{Three-dimensional entanglement-based (3DEB) protocol}

Let us now assume that the source emits the maximally entangled qutrit state
$\ket{\phi_3^+}$ and that Alice and Bob share this entangled pair and perform
measurements along one of the four optimal bases described above. It is
easy to check that $\ket{\phi_3^+}$ may be rewritten as
\begin{equation}
\ket{\phi_3^+}={1\over\sqrt{3}}(\ket{0_{\phi}}\otimes
\ket{0_{\phi}^*}+\ket{1_{\phi}}\otimes
\ket{1_{\phi}^*}+\ket{2_{\phi}}\otimes
\ket{2_{\phi}^*},
\end{equation}
 where
\begin{equation}
\ket{l_{\phi}^*}={1\over
\sqrt 3}\sum_{k=0}^2e^{-ik({2\pi l\over 3}+\phi)}\ket{k} \qquad (l=0,1,2).
\end{equation} Therefore, when Alice performs a measurement in the $\phi$
basis $\{\ket{l_{\phi}}\}$ and Bob in the conjugate basis
$\{\ket{l_{\phi}^*}\}$, their results are 100\% correlated. 
In addition, the four optimal bases defined above
can be shown to be 100\% correlated two by two. This can be
understood by noting that phase conjugation corresponds to a symmetry
that interchanges the bases of the dodecagon.
\par

It is therefore natural to consider the following generalization of the
Ekert91 protocol for qutrits, which we shall denote the 3-dimensional
entangled-based (3DEB) protocol\cite{protoc}. In this protocol, Alice and
Bob share the entangled state $\ket{\phi_3^+}$ and choose each their
measurement basis at random among one of the four bases maximizing
violation of local realism (according to the statistical distribution that
they consider to be optimal). Because of the existence of 100\% correlations
between measurements in local bases of the same $\phi_i$, a fraction of the
measurement outcomes can be used in order to establish a deterministic
cryptographic key. The rest of the data, for the cases when the left
and right phases are different, can be used in order to detect 
the presence of an eavesdropper for example with the
of Bell inequalities of Ref. \cite{collins} or with the computer algorithm 
of Ref. \cite{zukozeil}. Let us now study the security of
this protocol against optimal individual attacks.

\section{Individual attacks and optimal qutrit cloning machines}

We use a general class of cloning transformations as defined in
\cite{CERFPRL,CERFA,CERF}. If Alice sends the input state
$\ket{\psi}$ belonging to an $N$-dimensional space (we will consider
$N=3$ later on), the resulting joint state of the two clones (noted $A$ and
$B$) and of the cloning machine (noted $C$) is
\beq  \label{transfo}
\ket{\psi}  \to
\sum_{m,n=0}^{N-1} a_{m,n} \; U_{m,n}\ket{\psi}_A \ket{B_{m,-n}}_{B,C} =
\sum_{m,n=0}^{N-1} b_{m,n} \; U_{m,n}\ket{\psi}_B \ket{B_{m,-n}}_{A,C},
\eeq where
 \beq
 U_{m,n}=\sum_{k=0}^{N-1} e^{2\pi i (kn/N)} \ket{k+m}\bra{k},
 \eeq and
\beq
\ket{B_{m,n}}=N^{-1/2} \sum_{k=0}^{N-1} e^{2\pi i (kn/N)} \ket{k}\ket{k+m},
\label{BELLST}
\eeq with $0\le m,n \le N-1$.
 $U_{m,n}$ is an ``error'' operator: it shifts the state by $m$ units (modulo
$N$)
 in the computational basis, and multiplies it by a phase so as to
 shift its Fourier transform by $n$ units (modulo $N$). The equation
(\ref{BELLST})
 defines the $N^2$ generalized Bell states for a pair of N-dimensional
systems.

Tracing over systems
$B$ and $C$ (or $A$ and $C$) yields the final states of clone $A$ (or clone
$B$): if the input state is $\ket{\psi}$, the clones $A$ and $B$ are in a
mixture of the states $\ket{\psi_{m,n}}=U_{m,n}\ket{\psi}$ with respective
weights $p_{m,n}$ and $q_{m,n}$:
\beq
\rho_A=\sum_{m,n=0}^{N-1} p_{m,n} \proj{\psi_{m,n}},
\qquad  \rho_B=\sum_{m,n=0}^{N-1} q_{m,n} \proj{\psi_{m,n}}
\eeq In addition, the weight functions of the two clones ($p_{m,n}$ and
$q_{m,n}$) are related by
\beq
 p_{m,n} = |a_{m,n}|^2,  \qquad
 q_{m,n} = |b_{m,n}|^2,
\eeq where $a_{m,n}$ and $b_{m,n}$ are two (complex) amplitude functions that
are dual under a Fourier transform \cite{CERFA,CERF}:
\beq  \label{FT}
 b_{m,n} = {1\over N} \sum_{x,y=0}^{N-1} e^{2\pi i {nx-my\over N}} a_{x,y}.
\eeq

Let us now  analyze the possibility of using such a cloning procedure in the
eavesdropping attack on the two entangled qutrit protocol. Therefore we put
$N=3$. Assume that Eve clones the state of the qutrit that is sent to Bob
(represented as the ket $\ket{\psi}$ in Eq.~\ref{transfo}), and resends the
imperfect clone (labeled by $A$) to Bob while she conserves the other one
(labeled by $B$). Then, in analogy with
\cite{boure2}, Eve will measure her clone in the same basis as Bob (the
$\phi$ basis) and her ancilla (labeled by $C$) in the conjugate basis (the
$\phi^*$ basis). For deriving Eve's information, we need first to rewrite the
cloning transformation in these bases. By straightforward computations we
get, when $\phi$ is equal to zero, that: \beq
\ket{B_{m,n}}=3^{-1/2} \sum_{l=0}^{2} e^{im({2\pi\over 3}(l-n)+\phi) }
\ket{l_{\phi}}\ket{(l-n)_{\phi}^*} =e^{im({-2\pi\over 3}n+\phi)} \ket{\tilde
B_{-n_{\phi},m_{\phi}^*}},\eeq  where, by definition,
\beq
\ket{\tilde B_{m_{\phi},n_{\phi}^*}}=3^{-1/2} \sum_{k=0}^{2} e^{2\pi i (kn/3)}
\ket{k_{\phi}}\ket{(k+m)_{\phi}^*},
\eeq  and \beq
 U_{m,n}=\sum_{k=0}^{2} e^{-im({2\pi\over 3}(k+n)+\phi)
}\ket{(k+n)_{\phi}}\bra{k_{\phi}}
 =e^{-im({2\pi\over 3}n+\phi)}\tilde U_{n_{\phi},-m_{\phi}},
\eeq where the tilde subscript refers to the new ($\phi$ and $\phi^*$) bases.
After substitution in Eq.~\ref{transfo}, we get:
\beq  \label{transfophi}
\ket{\psi}  \to
\sum_{m,n=0}^{2} a_{m,n} \; U_{m,n}\ket{\psi}_A \ket{B_{m,-n}}_{B,C}
 =\sum_{m,n=0}^{2}\tilde a_{m,n} \;
\tilde U_{m_{\phi},n_{\phi}}\ket{\psi}_A \ket{\tilde
B_{m_{\phi},n_{\phi}}}_{B,C},
\eeq where the new amplitudes are defined as $\tilde a_{n,-m}=a_{m,n}$.
\par

We are interested in a cloning machine that has the same effect when
expressed in the four optimal bases, i.e. when
$\phi_i\,=\,{2\pi\over 12}\cdot i (i=0,1,2,3)$. This imposes strong
constraints on the amplitudes
$a_{m,n}$ characterizing the cloner, which must be of the form
\beq \label{ampl} (a_{m,n})=
\left( \begin{array}{ccc} v & x & x \\
 y & y & y \\
 z & z  & z
\end{array} \right).
\eeq
It is possible to check that, in analogy with the qubit case \cite{bruss},
such a cloner is phase-covariant, which means that it acts identically on
each state of the $\phi$-bases. In particular, the identity (\ref{transfophi})
can be shown to hold for all values of $\phi$. The reason for this
property is that, roughly speaking, if the cloner remains invariant when
expressed in several bases, then it means that certain combinations of Bell
states possess several Schmidt bi-orthogonal decompositions. It is well-known
that when at least two such decompositions exist for a bipartite pure state,
then there exist infinitely many. This explains why requiring the same
cloning fidelity in two optimal bases ($\phi_i\,=\,{2\pi i\over 12},
\phi_j\,=\,{2\pi j\over 12}$ with $i,j=0,1,2,3$ and $i\ne j$) implies
phase-covariance (i.e., $\phi$ arbitrary). A proof of this property is out of
the scope of the present paper.
\par

Let us now evaluate the fidelity of this phase-covariant cloner for qutrits,
along with the information that Bob and Eve obtain about Alice's state. The
fidelity of the first clone (the one that is sent to Bob) when copying a
state $\ket{\psi}$ can be written, in general, as
\beq \label{fidelity} F_A=\langle\psi|\rho_A|\psi\rangle = \sum_{m,n=0}^{N-1}
|a_{m,n}|^2  |\langle\psi|\psi_{m,n}\rangle|^2.
\eeq Of course, the same relation holds for the second clone (the one that is
kept by Eve) by replacing $a_{m,n}$ by $b_{m,n}$. For the cloning machine
defined by Eq.(\ref{ampl}), it is possible to compute the fidelities when
cloning the component states of the $\psi$-bases by a straightforward but
lengthy computation. It can be shown that the fidelity of the first clone
does not depend on $\phi$, that is,
\beq F_A=\langle l_{\phi}|\rho_A|l_{\phi}\rangle=v^2+y^2+z^2
\eeq for all $\phi$. The disturbances $D_{A1}$ and $D_{A2}$ of the first
clone, defined respectively as $\langle l_{\phi+{2\pi\over
3}}|\rho_A|l_{\phi+{2\pi\over 3}}\rangle$ and
$\langle l_{\phi-{2\pi\over 3}}|\rho_A|l_{\phi-{2\pi\over 3}}\rangle$ yield
both $x^2+y^2+z^2$. Making use of Eq.~(\ref{FT}), we obtain that, for the
second clone, the states of the  bases used in the cryptographic protocol are
all copied with the same fidelity, which is maximum when $y=z$, and is given
by
\beq F_B = (v^2+2x^2+12y^2+8xy+4vy)/3.
\eeq Also, we get the same disturbance for all $\phi$ (minimal when $y=z$)
given by $ D_{B1}=D_{B2}=(v^2+2x^2+3y^2-4xy-2vy)/3$.
\par

We must now find what is the optimal strategy for Eve. In virtue of the
phase-covariance and in order to simplify the notations, we shall from now on
omit the labels that refer to the particular basis $\phi$ in which the
measurement is carried out. After substitution in Eq.~(\ref{transfo}), we get
\beq
\ket{\psi_k} \to 3^{-{1\over 2}}\sum_{m,l=0}^{2}
\tilde c_{m,k-l} \;  \ket{\psi_{k+m}}_A
\ket{\psi_{l}}_B\;\ket{\psi_{l+m}}_C,
 \eeq where $\tilde c_{m,j}=\sum_{n=0}^{2}\tilde a_{m,n}e^{i{2\pi\over 3}jn}$.
Now, $\tilde
a_{m,n}=y+\delta_{n0}((v-y)\delta_{m0}+(x-y)(\delta_{m1}+\delta_{m2}))$ so
that
$\tilde
c_{m,j}=(3y\delta_{j0}+(v-y)\delta_{m0}+(x-y)(\delta_{m1}+\delta_{m2}))$.
Therefore,
\beqa
\ket{\psi_k} \to   3^{-{1\over
2}}\{\ket{\psi_k}_A(3y\ket{\psi_{k}}_B\ket{\psi_{k}}_C+(v-y)\sum_{l=0}^{2}
\ket{\psi_{l}}_B\ket{\psi_{l}}_C)+
\nonumber \\
\ket{\psi_{k+1}}_A(3y\ket{\psi_{k}}_B\ket{\psi_{k+1}}_C+(x-y)\sum_{l=0}^{2}\ket{
\psi_{l}}_B\ket{\psi_{l+1}}_C)+
\nonumber \\
\ket{\psi_{k-1}}_A(3y\ket{\psi_{k}}_B\ket{\psi_{k-1}}_C+(x-y)\sum_{l=0}^{2}\ket{
\psi_{l}}_B\ket{\psi_{l-1}}_C)\}.
\eeqa After Alice's (or Bob's) measurement basis is disclosed, Eve's optimal
strategy can be shown \cite{boure2} to be the following: first she measures
both her copy $B$ and the cloning machine $C$ in the same basis as Bob, the
difference (modulo 3) of the outcomes simply giving Bob's error $m$.
Conditionally on Eve's measured value of $m$ (i.e., conditionally on Bob's
error), the information Eve has on the state $\ket{\psi}$ can be expressed as
\begin{eqnarray}
I(A{\rm:}E|m = 0)  &=& \log(3) - H\left[\frac{(v +
2y)^2}{3F_A}, \frac{(v - y)^2}{3F_A},\frac{(v - y)^2}{3F_A}\right]
\nonumber \\
I(A{\rm:}E|m \neq 0)  &=& \log(3) - H\left[\frac{2(x+2y)^2}{3(1-F_A)},
\frac{2(x-y)^2}{3(1-F_A)},\frac{2(x-y)^2}{3(1-F_A)}\right],
\end{eqnarray}
where $F_A=v^2+2y^2$ since we have $y=z$. On average, we get
for Eve's information
\beq
I_{AE}=F_A \; I(A{\rm:}E|m = 0)+(1-F_A)\; I(A{\rm:}E|m \neq 0).
\eeq
Of course, Bob's information is given by
\beq
I_{AB}= \log(3) - H\left[F_A, \frac{1-F_A}{2},\frac{1-F_A}{2}\right].
\eeq
We now use a theorem due to Csisz\'{a}r and K\"{o}rner \cite{Csiszar}
which provides a lower bound on the secret key rate, that is, the rate $R$ at
which Alice and Bob can generate secret key bits via privacy amplification:
if Alice, Bob and Eve share many independent realizations of a probability
distribution $p(a,b,e)$, then there exists a protocol that generates a number
of key bits per realization satisfying
\beq  \label{csiszar}
R \geq \max(I_{AB}-I_{AE},I_{AB}-I_{BE})
\eeq
In our case, $I_{AE}=I_{BE}$ since Eve knows exactly Bob's error $m$. It
is therefore sufficient that $I_{AB}>I_{AE}$ in order to establish a secret
key with a non-zero rate. If we restrict ourselves to one-way communication
on the classical channel, this actually is also a necessary condition.
Consequently, the quantum cryptographic protocol above ceases to generate
secret key bits precisely at the point where Eve's information matches Bob's
information.
\par

We thus need to estimate the maximal fidelity $F_A$ (or minimal error rate)
for which a cloning machine exists such that $I_{AE}=I_{AB}$. This
constrained optimization problem can be solved numerically, giving
\beq   \label{F_A} F_A=0.7753
\eeq corresponding to the solution
$(v,x,y)=(0.8320,0.1711,0.2038)$. Since $x\ne y$, this optimal cloner is
therefore distinct from the universal qutrit cloner (which clones all states
with the same fidelity). Actually, it is slightly better than
the (asymmetric) universal qutrit cloner, which gives a fidelity 
$F_A = 0.7733 $
at the crossing point of Bob's and Eve's information curves \cite{boure2}.
This means that the quantum cryptographic protocol where the four mutually
unbiased qutrit bases are used (see \cite{HBech-APeres}) 
is slightly better than the 3DEB protocol 
as it admits a $0.2\%$ higher error rate 
($1-F_A=22.67\%$ instead of $22.47\%$).
\par

The cloner that we have derived here
is an asymmetric version of the so-called two-phase-covariant qutrit
cloner that is described in \cite{Cloning-a-qutrit,presti} [this symmetric
two-phase-covariant qutrit cloner has a fidelity
$(5+\sqrt{17})/12\approx 0,760$].  It copies all states of the form
$3^{-1/2}(\ket{0}+e^{i\alpha}\ket{1}+e^{i\beta}\ket{2})$ with a fidelity
0.7753 ($>0.7733$) for all $\alpha$ and $\beta$,
while the states of the computational basis $\{\ket{0},
\ket{1},\ket{2}\}$ are cloned with a lower fidelity 0.7507 ($<0.7733$).
Actually, its relation with the symmetric two-phase-covariant cloner is of
the same kind as the relation
between the asymmetric universal qutrit cloner (of fidelity $0.7733$)
and the symmetric universal qutrit cloner (of fidelity ${3\over 4}$).
\par

\section{Conclusions}

The Ekert91 protocol and its qutrit extension, the 3DEB protocol which is
analyzed in the present paper, involve encryption bases for which the
violation of local realism is maximal. If Alice and Bob measure their member
of a maximally-entangled qutrit pair in two ``conjugate'' bases, this gives
rise to perfect correlations. After  measurement is performed on each member
of a sequence of maximally-entangled qutrit pairs, Alice and Bob can reveal
on a public channel what were their respective choices of basis and identify
which trit was correctly distributed, from which they will make the key.
They can use the rest of the data in order to check that it does not admit a
local realistic simulation. For instance they can check that their
correlations violate some generalized Bell or CHSH inequalities. As the
resistance of such a violation against noise is maximal when the
maximally-entangled qutrit pair is measured in the optimal qutrit bases
discussed here
(and is higher than all what can be achieved with qubits), the 3DEB protocol
is optimal from the point of view of the survival of non-classical
correlations in a noisy environment.
\par

Indeed, our results imply that the 3DEB protocol
is more robust against optimal incoherent attacks than the
Ekert91 qubit protocol. This is because 
the optimal qubit phase-covariant cloning
machine (which clones the optimal qubit bases involved in CHSH
with the same fidelity) gives a somewhat
higher fidelity $F_A={1\over 2}+{1\over\sqrt{8}} \simeq 0.8536$
\cite{bruss,Cloning-a-qutrit,HBech-NGisin}
than Eq.~(\ref{F_A}). 
In other words, the acceptable error rate, i. e. the error rate
$1-F_A$ above which the security against incoherent attacks is not
ensured, is 22.47\% for the 3DEB protocol, while it is only 14.64\% for
Ekert91.
\par

Recently, it has been shown that the violation of a Bell inequality extended
to qutrits is possible, as long as
 the ``visibility" of the two-qutrit interference exceeds $V_{thr}={6\sqrt
3-9\over 2} \simeq 0.6962$
\cite{chenzuko,collins}. The visibility mentioned above is directly related the
threshold fraction of  unbiased noise, $(1-V_{thr})$, which has to be admixed
to the maximally entangled state in order to erase the non-classical
character of the correlations, and therefore is a measure of robustness of
such  a non-classicality \cite{zukozeil}. This means that the non-existence
of a local realistic model of the correlations is guaranteed if the fidelity
$F_A$ that characterizes the communication channel between Alice and Bob,
(detectors included, so 1-$F_A$ is the effective error rate in the
transmission) is larger than ${2\over 3}\times 0.6962 +{1\over 3}\approx
0.7974$ (instead of ${1\over 2}+{1\over\sqrt{8}} \simeq 0.8536$ in the case of
qubits \cite{cirelson,chsh,RMP}). On the other hand, we have shown here that
the 3DEB protocol is secure against a cloning-based individual attack, if
$F_A>0.7753$. Consequently, when a violation of a qutrit Bell inequality
\cite{chenzuko,collins} occurs, the security of the 3DEB protocol against
individual attacks is automatically guaranteed.  
Therefore, the violation of Bell
inequalities is a {\em sufficient} condition for security, as it implies
that Bob's fidelity is higher than the security threshold. Remarkably, for
qubits, the corresponding sufficient condition 
($F_A > 0.8536$)
is also necessary \cite{RMP} (this is
apparently the case for qubits only).
\par

In addition, the violation of Bell inequalities guarantees that the 3DEB
protocol is secure against so-called Trojan horse attacks during which the
eavesdropper would control the whole transmission line and replace the signal
by a fake, predetermined local-variable dependent, signal that mimics the
quantum correlations. Such an attack can be thwarted when the signal is
encrypted in the the optimal bases provided that the noise level is low
enough (including now also the inefficiency of the detectors) so that
no such local realistic simulation of the signal does exist, and provided
that Alice and Bob perform their respective choices of bases independently
and quickly enough \cite{Durttrojan} so that their measurements are
independent spatially-separated events. Note that all
the protocols in which mutually unbiased bases are involved but with no
entanglement (such as BB84\cite{BB84}, the 6-state qubit
protocol\cite{DBruss,HBech-NGisin}, or the 12-state qutrit
protocol\cite{HBech-APeres}) admit
a local realistic model, so that they are not secure against Trojan horse
attacks.
\par

Finally, it is interesting to compare the performances of the 3DEB protocol to
those of the 3-dimensional extension of BB84.
The cloner that must be used in the latter case,
where two mutually unbiased qutrit bases are used, has a fidelity
of 0.7887 \cite{boure2}, thus a bit higher than the fidelity of the
cloner analyzed here, see Eq. (\ref{F_A}). 
Therefore, the 3DEB protocol also gives a slightly
higher acceptable error rate than the 3-dimensional extension of BB84
($22.47\%$ instead of $21.13\%$).
This, together with the robustness with respect to Trojan horse attacks,
clearly establishes the advantage of entanglement-based
protocols with respect to BB84-like protocols.
\par

In summary, we have derived a  qutrit cloning machine that clones equally
well the four optimal qutrit bases (those which maximize the violation
of local realism), so it gives the optimal individual attack
in the 3DEB protocol introduced here.
The acceptable error rate of the 3DEB protocol 
turns out to be 22.47\%, which is higher than that
of Ekert91 qubit protocol 
(as well as that of the 3-dimensional extension of BB84).
Our analysis thus confirms a seemingly general property that qutrit schemes
for quantum key distribution are more robust against noise than the
corresponding qubit schemes.
\par

\medskip

{\it Note:} After completion of this work, an independent paper by
D. Kaszlikowski {\it et al.} has appeared\cite{dagomir}, which
shows that, if Eve acts on one member of a maximally entangled qutrit pair,
then her information attains Alice and Bob's mutual information
at a visibility of 0.6629. In our notation, this means that the fidelity
at the information crossing point is
${2\over 3}\times 0.6629 + {1\over 3}\simeq 0.7753$, which exactly coincides
with our Eq.~(\ref{F_A}). Nevertheless the two approaches are different in
the following sense: in our approach, we assume that Eve clones the state of
the qutrit that is sent to Bob according to Eq.~(\ref{transfo}) and then we
impose that the cloning fidelity is identical for all the states of the
$\phi$ bases in order to fix the parameters $a_{m,n}$. Instead,
in \cite{dagomir}, a general transformation is postulated from the beginning,
and extra-constraints are imposed. We have also checked that
our optimal cloning machine satisfies these constraints, so the two
approcahes are compatible. Our approach being constructive, we
obtain the explicit form of the cloner, which is not the case in the approach
of \cite{dagomir}. Moreover, although the optimal cloning machines coincide
in both approaches, it can be
shown that our approach allows us to build new and more general solutions that
satisfies the constraints considered in \cite{dagomir}.
\par
\medskip

\leftline{\large \bf Acknowledgment}
\medskip

We are grateful to J. Roland for carrying out the numerical calculations. We
thank the European Science Foundation for financial support. T.D. is a
Postdoctoral Fellow of the Fonds voor Wetenschappelijke Onderzoek,
Vlaanderen. N.J.C. and N.G. acknowledge funding by the
European Union under the project EQUIP (IST-FET programme). M.\.Z.
acknowledges the KBN grant No. 5 P03B 088 20.

\end{document}